\documentclass[showpacs,preprintnumbers,amsmath,amssymb]{revtex4}
\usepackage{graphicx}
\begin{document}
\preprint{IMAFF-RCA-06-09}
\title{Accelerating Hilbert-Einstein universe without dynamic dark energy}

\author{Pedro F. Gonz\'{a}lez-D\'{\i}az and Alberto Rozas Fern\'{a}ndez}
\affiliation{Colina de los Chopos, Centro de F\'{\i}sica ``Miguel A.
Catal\'{a}n'', Instituto de Matem\'{a}ticas y F\'{\i}sica
Fundamental,\\ Consejo Superior de Investigaciones Cient\'{\i}ficas,
Serrano 121, 28006 Madrid (SPAIN).}
\date{\today}
\begin{abstract}
By using an unmodified Einstein gravity theory it is shown that all
of the speeding-up effects taking place in the current universe are
entirely due to the quantum effects associated with the background
radiation or to the combination of such effects with those derived
from the presence of a cosmological constant, without invoking any
dynamic dark energy component. We obtain that in both cases the
universe accelerates at a rate slightly beyond what is predicted by
a cosmological constant but does not induce any big rip singularity
in the finite future.
\end{abstract}


\maketitle


\pagebreak

Nowadays theoretical cosmology appears to confront a rather puzzling
situation. Whereas current observations seem to point to less than
-1 values for the most probable estimates of the parameter $w$ of
the equation of state [1], the most popular theoretical quintessence
models are plagued with violent instabilities, violations of the
energy conditions, future singularities, so as unphysical scalar
field negative kinetic terms, or ghosts, for $w<-1$ [2]. Exorcizing
procedures have been therefore considered [3] in order to justify
why values $w<-1$ can be compatible with usual quintessence models,
or alikes, that actually correspond to greater than or equal to -1
$w$-parameters. None of such procedures have however been successful
so far [4]. Time-dependent equations of state associated with
tracked quintessence scenarios [5] have also been analyzed to solve
the above puzzle but they turned out to fail, too. Even more
unsuccessful have been several cosmological models based on the idea
that dark energy is not necessary to predict cosmic acceleration.
Included among these approaches are, on the one hand, those models
using modified theories of gravity whose action integral contains
extra terms to the familiar Hilbert-Einstein action [6] and, on the
other hand, some descriptions in which late-time acceleration could
be explained by inhomogeneities produced during primordial inflation
[7]. The first kind of such scenarios is in turn plagued with
theoretical inconsistencies and instabilities, and cannot be
accommodated with cosmological observations and solar physics
experiments [8]. The second type of these scenarios is most
interesting in that, besides avoiding the concourse of any
mysterious dark energy fluid or field, they do not invoke any
modifications of gravity and hence become most economical.
Unfortunately, it has been shown [9] that to second order in spatial
gradients, the corrections are unable to account for the observed
speed-up of the cosmic expansion.

In this letter we shall look at current acceleration by using the
same general economical philosophy as in the last kind of the
above models, even without invoking, moreover, any effects induced
at the primordial inflationary period. The sole ingredients which
we shall explicitly include, besides general relativity, are the
quantum effects on the trajectories of the particles that make up
the background radiation. Such effects will be modeled through the
relativistic generalization of the original sub-quantum potential
formalism by Bohm [10] and lead by themselves to an accelerating
expansion which, consistently, goes beyond what is predicted by a
cosmological constant. Thus, we use a version of the sub-quantum
model for dark energy [11] which will in principle be motivated by
an up-grading-to-scalar field method stemming from the analogy
with the classically-interpreted Hamilton-Jacobi equation derived
from the Klein-Gordon wave equation for a quasi classical wave
function $\Psi=R\exp(iS/\hbar)$, i.e.
\begin{equation}
E^2-p(v)^2+\tilde{V}_{SQ}^2=m_0^2 ,
\end{equation}
where
\begin{equation}
\tilde{V}_{SQ}=\hbar\sqrt{\frac{\nabla^2 R-\ddot{R}}{R}}
\end{equation}
is the sub-quantum potential, $v=\dot{q}(t)$ and $p=\partial
\tilde{L}/\partial\dot{q}$, with $\dot{}=d/dt$ and $\tilde{L}$ being
the Lagrangian
\begin{equation}
\tilde{L}=\int d\dot{q} p =\int
dv\sqrt{\tilde{V}_{SQ}^2-m_0^2+\frac{m_0^2}{1-v^2}} .
\end{equation}
As shown first by Bagla, Jassal and Padmanabhan [12] for the fully
classical case and later on by one of the present authors [11] for
the case that the Lagrangian contains a sub-quantum potential,
upgrading the quantities entering this simple Lagrangian to their
field-theory counterparts actually leads to a cosmological
tachyonic model which can be used to predict cosmic acceleration.
In order to tentatively motivate our cosmic model, following Ref.
[11], we shall replace then the quantity $q$ for a scalar field
$\phi$, the quantity $\dot{q}^2\equiv v^2$ for
$\partial_i\phi\partial^i\phi\equiv \dot{\phi}^2$ and the rest
mass $m_0$ for the potential $\tilde{V}(\phi)$. With these
replacements and leaving $\tilde{V}_{SQ}$ constant for the moment,
we can then integrate Eq. (3) to have for the field Lagrangian
$\tilde{L}=-\tilde{V}(\phi)E\left(x(\phi),k(\phi)\right)$, with
$E(x,k)$ the elliptic integral of the second kind,
$x(\phi)=\arcsin\sqrt{1-\dot{\phi}^2}$ and $k=
\sqrt{1-\tilde{V}_{SQ}^2/\tilde{V}(\phi)^2}$. At first sight one
should also up-grade $\tilde{V}_{SQ}$ to depend on $\phi$.
However, it will be seen later that such a up-grading would lead
to a final expression for $\tilde{V}_{SQ}$ which depends only on
$\dot{\phi}$, a dependence that disappears because for the present
model it is necessary that $\dot{\phi}$ be constant in order to
avoid divergences.

Even though the up-grading-to-field method has been so far used to
just suitably motivate the introduction of a cosmic field model,
such a method will be in the present scenario shown to be more
than a mere motivating procedure devoid of any physical
significance [13]. Actually, even after up-grading, the above
model can still be interpreted as physically describing pure
background radiation equipped with a sub-quantum potential, taking
dark energy to be nothing but the effect left in the classical
universe by that sub-quantum potential, provided the following two
conditions are fulfilled by the field theory that results after
up-grading: (1) the field potential $\tilde{V}(\phi)$ is
identically equal to zero, and (2) the time derivative of the
scalar field becomes $\dot{\phi}^2=1$. In fact, since the
sub-quantum potential has not been up-graded to any
field-depending quantity, if such two conditions are either shown
to hold or imposed, then the up-grading process can readily be
seen to be equivalent to a identity operation, leaving the
original particle theory essentially unchanged; i.e. the radiation
particles and the sub-quantum potential can also be regarded as
the unique physically relevant ingredients for the model. We note
that in a FRW framework $\dot{\phi}^2=1$ necessarily implies
$\phi=q$, and hence the second condition amounts to
$V(\phi)=V(q)=m_0=0$, so that, restoring the speed of light as
$c$, we have $\dot{q}^2=v^2=c^2$. In any event, in what follows we
shall eliminate any trace of all classical quantities from our
model, thereby representing dark energy by solely the sub-quantum
potential, a hidden quantity that has not been up-graded and that
by itself should necessarily be associated with the particles
described by Lagrangian (3), not with any field quantity. Thus,
the resulting dark-energy scenario would not have any classical
analog. It follows that the condition that we have to impose to
the scalar field theory derived in the sub-quantum model [11] to
satisfy the requirement that dark energy disappears once we erase
any trace of the background quantum effects is that the
Lagrangian, energy density and pressure turn all out to only
depend on the sub-quantum potential and will all vanish in the
limit where any possible cosmological constant and the sub-quantum
potential are both zero, i.e. $\Lambda\rightarrow 0$,
$\tilde{V}_{SQ}\rightarrow 0$. It will be seen in what follows
that the above conditions are all fulfilled provided that we start
with a Lagrangian density given by
\begin{equation}
L=-V\left(E(x,k)-\sqrt{1-\dot{\phi}^2}\right),
\end{equation}
where again $x=\arcsin\sqrt{1-\dot{\phi}^2}$ and now
$k=\sqrt{1-V_{SQ}^2/V^2}$, with $V\equiv V(\phi)$ the density of
potential energy associated to the field $\phi$. We do not expect
$\tilde{V}_{SQ}$ to remain constant along the universal expansion
but to increase like the volume of the universe ${\rm V}$ does. It
is the sub-quantum potential density $V_{SQ}=\tilde{V}_{SQ}/{\rm V}$
appearing in Eq. (4) what should be expected to remain constant at
all cosmic times. In fact, from the imaginary part of the
Klein-Gordon equation applied to the wave function $\Psi$ we can get
$v.\nabla R-\dot{R}$ and hence the continuity equation for the
probability flux $J=\hbar\; {\rm Im}(\Psi^*\nabla\Psi)/(m{\rm V})$,
$\nabla . J-\dot{{\rm P}}=0$, where ${\rm P}$ is the probability
density ${\rm P}={\rm Probability}/{\rm V}$. This continuity
equation is the mathematical equivalent of a probability
conservation law. Up-grading then the velocity $v$ to $\dot{\phi}$
and noting that $\dot{\phi}=\pm 1$ (see later) it follows that
$(\nabla^2 R-\ddot{R})/R=(\nabla^2 P-\ddot{P})/(2P)$, with $P=R^2$.
Assuming that the particles move locally according to some causal
law [10], one can now average Eq. (1) with the probability weighting
function $P=R^2$, so that one obtains for the averaged sub-quantum
potential squared, $\langle\tilde{V}_{SQ}^2\rangle_{av}=\int\int\int
dx^3 P\tilde{V}_{SQ}^2 =\hbar^2\int\int\int dx^3(\nabla^2 P
-\ddot{P})\equiv \hbar^2 \left(\langle\nabla^2
P\rangle_{av}-\langle\ddot{P}\rangle_{av}\right)$. Since the
universe is isotropic and homogeneous, the corresponding cosmic
conserved quantity can then be obtained by simply taking
$\langle\tilde{V}_{SQ}^2\rangle_{av}^{1/2}/{\rm V}=\langle
V_{SQ}^2\rangle_{av}^{1/2}$, that is, renaming for the aim of
simplicity all the quantities $\langle f^2\rangle_{av}^{1/2}$
involved in the averaged version of Eq. (1) as $f$, we can again
derive Eq. (4), now with $V_{SQ}$ a constant conserved quantity when
referred to the whole volume ${\rm V}$ of the isotropic and
homogeneous universe.

It is easy to see that in the limit of vanishing $V_{SQ}$, $VE(x,k)$
reduces to $\sqrt{1-\dot{\phi}^2}$ so that the Lagrangian (4)
vanishes as required. The pressure and energy density are then
obtained from Eq. (4) to read
\begin{equation}
p_{\phi}=-V\left(E(x,k)-\sqrt{1-\dot{\phi}^2}\right)
\end{equation}
\begin{equation}
\rho_{\phi}=V\left(\frac{\sqrt{\dot{\phi}^2+\frac{V_{SQ}^2}{V^2}(1-
\dot{\phi}^2)}\dot{\phi}}{\sqrt{1-\dot{\phi}^2}}
+E(x,k)-\frac{1}{\sqrt{1-\dot{\phi}^2}}\right) ,
\end{equation}
where we have considered $V\equiv V(\phi)$. In any case, for a
source with parameter $w(t)=p_{\phi}/\rho_{\phi}$ we must always
have
\begin{equation}
\frac{\dot{\rho}_{\phi}}{\rho_{\phi}}=-3H\left(1+w(t)\right)=\frac{2\dot{H}}{H}
.
\end{equation}
By itself this expression can generally determine the solution for
the scale factor $a(t)$, provided $w$= const. In such a case, we
obtain after integrating Eq. (7) for the scale factor
\[a=\left(a_0^{3(1+w_0)/2}+\frac{3}{2}(1+w_0)\kappa t\right)^{2/[3(1+w_0)]},\]
in which $a_0$ is the initial value of the scale factor and $\kappa$
is a constant. However, we shall not restrict ourselves in this
letter to a constant value for the parameter $w$ of the equation of
state but leave it as a time-dependent parameter whose precise
expression will be determined later on. Combining now Eq. (7) with
the expression for $w(t)$ we can then obtain an expression for
$d(H^{-1})/dt$ by using Eqs. (5) and (6) as well. Moreover,
multiplying Eqs. (5) and(6) and using Eq. (7), a relation between
the potential density $V$ and the elliptic integral $E$ can be
derived from the Friedmann equation $H^2= 8\pi G\rho_{\phi}/3$.
These manipulations allow us to finally obtain
\begin{widetext}
\begin{equation}
E=-\left[\frac{A(\dot{\phi},V,V_{SQ})\dot{\phi}\left(1+\frac{3H^2}{2\dot{H}}\right)
-1-\frac{3H^2\dot{\phi}^2}{2\dot{H}}}{\sqrt{1- \dot{\phi}^2}}\right]
=-\left\{\frac{\frac{3H^2\dot{\phi}^4
V_{SQ}^2}{\dot{H}}-\left(\frac{2\dot{H}}{G}\right)^2
(1-\dot{\phi}^2) +\dot{\phi}^2 V_{SQ}^2
(1+\dot{\phi}^2)}{\sqrt{1-\dot{\phi}^2}\left[\left(\frac{\dot{H}}{4\pi
G}\right)^2 -\dot{\phi}^2 V_{SQ}^2\right]}\right\}
\end{equation}
\end{widetext}
with
$A(\dot{\phi},V,V_{SQ})=\sqrt{\dot{\phi}^2+\frac{V_{SQ}^2}{V^2}(1-
\dot{\phi}^2)}$, and
\begin{equation}
V=-\frac{2\pi
G\sqrt{1-\dot{\phi}^2}}{\dot{H}\dot{\phi}^2}\left[\left(\frac{\dot{H}}{4\pi
G}\right)^2-\dot{\phi}^2 V_{SQ}^2\right].
\end{equation}

Thus, simple general expressions for the energy density and pressure
can be finally derived to be
\begin{equation}
\rho_{\phi}= 6\pi G\left(\dot{H}^{-1}H\dot{\phi} V_{SQ}\right)^2
\end{equation}
\begin{equation}
p_{\phi}=-4\pi G\dot{H}^{-1}\dot{\phi}^2 V_{SQ}^2
\left(1+\frac{3H^2}{2\dot{H}}\right) =w(t)\rho_{\phi} ,
\end{equation}
where
\begin{equation}
w(t)=-\left(1+\frac{2\dot{H}}{3H^2}\right) .
\end{equation}

The Friedmann equation $H^2=8\pi G\rho_{\phi}/3$, derived from the
action integral with the Lagrangian (4), corresponds to a universe
dominated by sub-quantum energy. Using Eq. (10) this Friedmann
equation leads to
\begin{equation}
\dot{H}=\pm 4\pi G\dot{\phi}V_{SQ} ,
\end{equation}
with a slowly-varying $w(t)$ that should probably be quite close,
but still less than -1 (that is, the case that current
observations each time more clearly are pointing to [1]). We have
also
\begin{equation}
H=\pm 4\pi G\phi V_{SQ} + C_1 ,
\end{equation}
with $C_1$ an integration constant. Note that from Eqs. (9) and
(13) it follows that $V(\phi)=0$, which is just one of the two
conditions required to make consistent our interpretation.
Moreover, if we assume that $\dot{\phi}$ is constant (an
assumption which would indeed be demanded by the fact that $v^2=1$
for radiation), then from the equation of motion that corresponds
to the Lagrangian for the field $\phi$ alone [12]
$\ddot{\phi}+(1-\dot{\phi}^2)(3H\dot{\phi}+ dV/Vd\phi)=0$, we have
$\dot{\phi}^2=+1$. Actually, from the Lagrangian density
$L_{SQ}=-V(\phi)E(x,k)$ we can also obtain,
\begin{widetext}
\[\dot{\phi}\ddot{\phi}=(1-\dot{\phi}^2)\left\{-3H\left[\dot{\phi}^2
+\frac{V_{SQ}^2}{V(\phi)^2}(1-\dot{\phi}^2)\right]
+\sqrt{1-\dot{\phi}^2}\sqrt{\dot{\phi}^2
+\frac{V_{SQ}^2}{V(\phi)^2}(1-\dot{\phi}^2)}\frac{\partial
L_{SQ}}{V(\phi)\partial\phi}-\frac{\partial
V}{V\partial\phi}\dot{\phi}^3\right\} ,\]
\end{widetext}
from which we again derive the conclusion that $\ddot{\phi}=0$
implies $\dot{\phi}^2=1$. Indeed, the assumption that
$\dot{\phi}^2=1$ can be really regarded as a regularity
requirement for $\ddot{\phi}$ because if $\dot{\phi}^2\neq 1$ then
$\ddot{\phi}$ would necessarily diverge since $V(\phi)$ vanishes
even when $\dot{\phi}^2\neq 1$, as it can be checked from Eqs. (9)
and (13). The same result can then be obtained from the equation
of motion derived from Lagrangian (4). Hence a vanishing
$\ddot{\phi}$ implies that strictly $\dot{\phi}^2=1$ and since in
addition $V=0$ the present model can be interpreted to describe
the cosmic sub-quantum effects necessarily associated with an
isotropic and homogeneous sea of bosonic particles with zero rest
mass which move at the speed of light, i.e. photons - identifying
that photon sea with the CMB is just a reasonable assumption. It
then follows that the condition $\dot{\phi}^2=\dot{q}^2=1$ becomes
a regularity requirement, and the condition $V(\phi)=V(q)=m_0=0$
results from the combined effect of the Friedmann equations and
the very nature of the model. We have now $\rho_{\phi}=\rho_q
=6\pi G(\dot{H}^{-1}HV_{SQ})^2=p_q/w(t)$ which in fact does not
depend on any field quantity, such as it was required for
interpreting dark energy as the sub-quantum energy associated with
radiation particles. The use of an up-grading-to-field motivating
method becomes thus rather superfluous in the present theory. We
had indeed obtained identical results and conclusions if we had
replaced $\phi$ and $V(\phi)$ for $q$ and $m_0$, respectively,
leaving $V_{SQ}$ unchanged, in Eqs. (4) - (14).

It follows then
\begin{equation}
H=\pm 4\pi GV_{SQ}t + C_0 ,
\end{equation}
in which $C_0$ is another integration constant, and for the scale
factor
\begin{equation}
a_{\pm}=a_0 e^{\pm 2\pi GV_{SQ}t^2 +C_0 t} .
\end{equation}
The solution $a_{-}$ would predict a universe which initially
expands but that immediately started to contract, tending to
vanish as $t\rightarrow\infty$. An always accelerating solution
slightly beyond the speeding-up predicted by a De Sitter universe
is given by the scale factor $a_{+}$. In what follows we shall
consider the latter solution as that representing the evolution of
our current universe and restrict ourselves to deal with that
solution only for the branch $t>0$, denoting $a_{+}\equiv a$ and
taking then $H$ and $\dot{H}$ to be definite positive.

Thus, the time-dependent parameter of the equation of state will be
given by
\begin{equation}
w(t)=-1-\frac{8\pi GV_{SQ}}{3(4\pi GV_{SQ}t+C_0)^2} ,
\end{equation}
which takes on values very close, though slightly less than -1 on
the regime considered so far.

Notice that in the limit $V_{SQ}\rightarrow 0$, $H$ becomes a
constant $H_0=C_0$, and hence $\rho_{\phi}\rightarrow 3C_0/(8\pi G)$
and $w\rightarrow -1$. Clearly, $H_0^2=\Lambda$ must be interpreted
as the cosmological constant associated with the De Sitter solution
$a=a_0 e^{H_0 t}$. When we set $C_0=0$ instead, then all remaining
quantities have the following limiting values
\begin{equation}
\rho_{\phi}=\frac{p_{\phi}}{w(t)}=6\pi GV_{SQ}^2 t^2 \rightarrow 0 ,
\end{equation}
\begin{equation}
w(t)= -1 -\frac{1}{6\pi GV_{SQ}t^2}\rightarrow -\infty
\end{equation}
and
\begin{equation}
a=a_0 e^{2\pi GV_{SQ}t^2}\rightarrow a_0 ,
\end{equation}
as $V_{SQ}\rightarrow 0$. That is precisely the result we wanted
to have and means that all the cosmic speed-up effects currently
observed in the universe can be attributed to the purely
sub-quantum dynamics that one can associate to the background
radiation, rather than to the presence of a dark energy component
or any modifications of Hilbert-Einstein gravity. In fact, it can
be readily checked that the expression obtained for $\dot{H}$
inexorably leads to a vanishing value for the potential $V(\phi)$,
and hence to $\dot{\phi}^2=1$, which correspond to pure radiation.
Consistency for the present theory is ensured by noticing that:
(i) $\dot{\phi}^2=1$ does clearly satisfy the Friedmann equation
$H^2=8\pi G\rho$, with $\rho=6\pi G(\dot{H}^{-1}HV_{SQ})^2$ and
that for the field $\phi$ from which that condition was derived,
and (ii) if we substitute $\dot{\phi}^2=\dot{q}^2=1$ and
$V(\phi)=V(q)=0$ back into Eqs. (5) and (6) and we use Eqs. (8)
and (9), we recover the regular values for energy density and
pressure given by Eqs. (10) and (11) for
$\dot{\phi}^2=\dot{q}^2=1$, which in fact show no dependence
whatsoever on any field quantity.

The result that, if there is not constant cosmological term, then
they are the considered sub-quantum effects associated with the
background radiation which are responsible for a current
accelerating expansion of the universe that goes beyond the
cosmological constant limit, implies, on the other hand, that (i)
the parameter of the equation of state is necessarily less than
-1, though probably very close to it, (ii) the energy density
increases with time, (iii) $\rho_{\phi}+p_{\phi}<0$, that is the
dominant energy condition is violated, and (iv) the kinetic term
of the equivalent field theory turns out to be $\dot{\phi}^2 >0$.
Whereas the first three properties are shared by the so called
phantom models [2], unlike such models, the fourth one guarantees
stability of the resulting universe because $V(\phi)=0$. Also
unlike the usual phantom scenarios, the present model does not
predict, moreover, any big rip singularity in the future. Finally,
the considered quantum effects may justify violation of the
dominant energy condition.

On the other hand, if we place a Schwarzschild black hole with
initial mass $M_0$ in the universe described by the suggested
model, the mechanism advanced by Babichev, Dokuchaev and Eroshenko
[14] would imply that the black hole will accrete this sub-quantum
phantom energy so that it would progressively lose mass down to
finally vanish at $t=\infty$, according to the equation
\begin{equation}
M=\frac{M_0}{1+\pi^2 DV_{SQ}M_0 t} ,
\end{equation}
with $D$ a constant. If we place a Morris-Thorne wormhole with
initial throat radius $b_0$ instead, the corresponding accretion
mechanism [15] leads now to a progressive increase of the wormhole
size governed by
\begin{equation}
b=\frac{b_0}{1-\pi^2 D' V_{SQ}b_0t},
\end{equation}
with $D'$ another constant, bringing us to consider the existence
of a big trip process [15] by which, relative to an asymptotic
observer at $r=\infty$, the wormhole will quickly grows up to
engulf the universe itself, blowing up at a finite time in the
future given by
\begin{equation}
\tilde{t}=\frac{1}{\pi^2 D' V_{SQ}b_0} .
\end{equation}
In this case, on times $t>\tilde{t}$ the wormhole converts into an
Einstein-Rosen bridge which decays into a black hole plus a white
hole that will in this case progressively lose mass to vanish at
$t=\infty$ [15]. This result holds both for a static wormhole
metric and when the throat radius is allowed to be time-dependent
[15].

Before closing up we shall briefly consider solution $a_-$. As it
has already been pointed out before, if
$C_0=H_0=\Lambda^{1/2}>>\sqrt{4\pi GV_{SQ}}$, then this solution
corresponds to an initial period of accelerating expansion with an
equation-of-state parameter $w$ greater, though very close to -1.
This situation would stand until a time
\begin{equation}
t_a= \frac{H_0-\sqrt{4\pi GV_{SQ}}}{4\pi GV_{SQ}},
\end{equation}
which corresponds to $w=-1/3$. After $t_a$ the universe would keep
expanding but now in a decelerating way until a time
\begin{equation}
t_c=\frac{H_0}{4\pi GV_{SQ}} ,
\end{equation}
after which the universe entered a contracting phase which would be
maintained until $t=\infty$. If $H_0 \leq \sqrt{4\pi GV_{SQ}}$, then
the present model would no longer be valid.

It could be at first sight thought that the universe might now be in
the phase $t<t_a$ of solution $a_-$, but current constraints on $w$
[1] seem to preclude that it can be greater than -1. Perhaps another
argument against solution $a_-$ be the fact that for this kind of
solution, while the accretion of the sub-quantum energy onto a
Morris-Thorne wormhole leads to a progressive decrease of the
wormhole size according to the law $b=b_0/(1+\pi^2 D' V_{SQ}t)$, the
size of a black hole of initial mass $M_0$ will progressively
increase with sub-quantum energy accretion so that $M=M_0 /(1-\pi^2
DV_{SQ}M_0 t)$. In this way, at a time $t_* =1/(\pi^2 DV_{SQ}M_0)$
the black hole would blow up. Clearly, for a supermassive black hole
at a galactic center one would then expect that by the present time
the black hole had grown up so big that its astronomical effects
would be probably observable.

All the above results have been obtained in the case that the energy
density associated with the sub-quantum potential would dominate
over any other type of energy. More realistic models where
contributions from dark and observable matters are taken into
account as well will be considered elsewhere.

\acknowledgements

\noindent The authors thank C. Sig\"{u}enza for useful discussions.
This work was supported by MCYT under Research Project No.
FIS2005-01181.

\end{document}